\documentclass{article}

\def\Ro{{\bf R}}
\def\Io{{\bf I}}
\def\Co{{\bf C}}
\def\avar{a}

\renewcommand{\Im}{\,{\rm Im}\,}
\renewcommand{\Re}{\,{\rm Re}\,}
\def\be{\begin{equation}}
\def\ee{\end{equation}}
\def\bea{\begin{eqnarray}}
\def\eea{\end{eqnarray}}

\begin{document}

\title{Covariance approach to the free photon field}

\author{Maciej Kuna\\
     Wydzia\l\ Fizyki Technicznej i Matematyki Stosowanej,\\
     Politechnika Gda\'{n}ska,\\
     ul.~Narutowicza 11/12, 80-952 Gda\'{n}sk, Poland\\
     E-mail: maciek@mifgate.mif.pg.gda.pl\\
     \\
     Jan Naudts\\
     Departement Natuurkunde, Universiteit Antwerpen UIA,\\
     Universiteitsplein 1, 2610 Antwerpen, Belgium\\
     E-mail: Jan.Naudts@ua.ac.be
}

\date{}
\maketitle

\begin{abstract}
We introduce photon theory following the same principles as for
introduction of the quantum theory of a single particle, using a
$C^*$-algebraic approach based on covariance systems. The basic
symmetries are additivity of the fields and additivity of test
functions. We write down in explicit form a state of this
covariance system. It turns out to reproduce the traditional
Fock representation of the free photon field, with a Lorentz
invariant vacuum. Properties of smeared-out photons are discussed.
\end{abstract}

\section{Introduction}

\section*{Motivation}

This paper is a first attempt to reformulate photon theory. It is
motivated by dissatisfaction with expositions in present day
textbooks. As Scharf\cite{SG89} notes, the fact that there are
various essentially different methods of quantizing the radiation
field shows that there are some difficulties with the subject. Two
problems must be recognized.

A first problem arises because of the use of the vector
potential $A_\mu(q)$, which is not uniquely determined by the
electromagnetic fields. The resulting
gauge freedoms can be tackled in many ways. Most often used is
the Gupta-Bleuler gauge, which is rather complicated to say the
least. Disadvantage of the radiation gauge is the lack of
manifest Lorentz covariance. Our treatment of the gauge problem
has been influenced by the work of Carey et
al\cite{CGH77}, which uses the Lorenz gauge. We show that
the photon states are invariant under the remaining
gauge freedom.

The next problem is that of positivity of the scalar product, in
combination with Lorentz invariance. Many textbooks abandon the
use of Hilbert spaces for this reason. Here, we give arguments
to restrict the set of classical wave functions. A side effect
is that the scalar product $\langle\psi|\phi\rangle$ of two
classical wave functions $\phi$ and $\psi$ satisfies the
positivity requirement. In addition, we prove explicitly in
Appendix D that the vacuum state does satisfy the positivity
condition. As a consequence, the quantum probabilistic
interpretation is saved in our approach.

\section*{Analogy}

The simplest example of a classical field is the vibrating
string. Its canonical variables are a displacement field
$\eta(q)$ and a conjugated momentum field $\pi(q)$. The correct
description of the quantized string can be obtained by a
limiting procedure starting from a chain of particles
interconnected with strings. Alternatively, one can describe the
quantized string as a covariance system\cite{NK00}
--- see Appendix A --- consisting of
an abelian algebra $\cal A$ generated by smeared-out displacement
fields, a group $G$, which is the group of adding fields, and the
obvious action of $G$ on $\cal A$. Such a description is very
analogous to the description\cite{NJ01} of standard quantum
mechanics as a covariance system, in which case $\cal A$ is an
algebra of functions of position, and $G$ is the group of shifts
in position. It is therefore natural to expect that also the
quantized electromagnetic field can be described as a covariance
system with additivity of fields as the basic symmetry group. The
first problem that one encounters when trying such an approach is
that the quantized smeared-out field operators $\hat A_\mu(\psi)$
do not form an abelian algebra. Indeed, in textbooks\cite {JR80}
one finds the commutation relations
\be
\left[\hat A_\mu(q),\hat A_\nu(q')\right]
=-ig_{\mu,\nu}D_0(q-q')
\label{(2.6)}
\ee
with the
Pauli-Jordan function $D_0(q)$ defined by
\be
D_0(q)={1\over(2\pi)^{3}}\int_{\Ro^3}\hbox{ d}{\bf k}
\,\exp\left(i\sum_{\alpha=1}^3{\bf k}_\alpha q_\alpha\right)
{1\over |{\bf k}|}\sin(q_0|{\bf k}|)
\label{(2.7)}
\ee
(we have
chosen the sign in (\ref{(2.6)}) in such a way that later on
creation and annihilation operators have their usual properties,
i.e.~the annihilation operator is complex linear in the field).
Smearing out (\ref {(2.6)}) with classical wave functions $\psi$
and $\phi$ gives (see Appendix B)
\bea
\left[\hat A(\psi),\hat A(\phi)\right]
&=&2i\Im\int_{\Ro^3}\hbox{ d}{\bf k}\,\frac{1}{2|{\bf k}|}
\overline{\psi^\mu({\bf k})}\phi_\mu({\bf k}).
\label{emfcr}
\eea
There are two possible interpretations of
these non-trivial commutation relations. It is a tradition in the
physics literature to interpret photons as excitations of a
harmonic oscillator. In this traditional approach\cite {JR80}, the
canonical variables to be quantized are the vector fields
$A_\mu(q)$ and their derivatives $\delta_\nu A_\mu(q)$. These
correspond with the displacement field $\eta(q)$ and conjugated
momentum field $\pi(q)$ of the harmonic string. By integration,
one then obtains (\ref {emfcr}). An alternative interpretation is
suggested by recent research on noncommutative spacetime. In the
latter context, spacetime positions $\hat Q_\mu$ satisfy
nontrivial commutation relations
\be
\left[\hat Q_\mu,\hat
Q_\nu\right]=i\epsilon_{\mu\nu}. \label{dopcom}
\ee
(see e.g.
Doplicher et al\cite {DFR94,DFR95} and Naudts and
Kuna\cite{NK01}). The $\hat Q_\mu$ are the generators of shifts in
momentum space. The commutation relations (\ref{dopcom}) can\cite
{NK01} be seen as a originating from a projective representation
of the group of shifts. By analogy, we can see the operators $\hat
A(\psi)$ as generators of the group of addition of classical wave
functions. In fact, it is clear\cite {NJ01} that the Weyl algebra
of the free photon field, as used e.g.~in Carey et
al\cite{{CGH77}}, can be replaced by a covariance system
$(\Co,H,\Io)$ consisting of the algebra $\Co$ of complex numbers,
the complex vector space of classical wave functions $H$, and the
trivial action $\Io$ of the latter on $\Co$.

\section*{Duality}

Because the addition of fields is the basic symmetry, rather
than addition of classical wave functions, it is obvious
to consider also the generators $\hat F(a)$ of the group
of adding fields
\be
a'({\bf k})\rightarrow a'({\bf k})+a({\bf k}).
\ee
Here, the fields are represented by Fourier coefficients $a_\mu({\bf k})$ of
the vector potential $A_\mu(q)$ (see next section).
In this context a duality between classical wave functions and
Fourier transformed vector potentials is of importance.
It implies a duality between the operators $\hat A(\psi)$
and $\hat F(a)$, similar to the duality between position
and momentum operators in standard quantum mechanics.

Technically speaking, there is no need to include this duality
in the formalism. Indeed, we will find that, in the Fock
representation, for each $a$ there exists a $\psi$ such that
$\hat F(a)=\hat A(\psi)$. However, this coincidence is
a special property of the photon field and might be absent in
more general field theories. For that reason we prefer to clarify
the dual r\^ole of the operators $\hat A(\psi)$ and $\hat F(a)$.

\section*{Notations}

We use Greek letters $\mu,\nu,\sigma,\cdots$ for indices that run from
0 to 3, in combination with Einstein's summing convention,
i.e., if such an index appears twice then a summation from 0 to 3
is understood. These Greek indices are lowered and raised
in the standard way, i.e., by definition is $x^\mu=g^{\mu\nu}x_\nu$,
with the metric tensor $g$ equal to the diagonal matrix
with eigenvalues $+1,-1,-1,-1$. The Greek index $\alpha$
will be used to label spatial components. Hence it runs from 1 to 3
and no summation convention is used for it. Vectors in $\Ro^3$
are written in boldface. Quite often, a four-vector $q$ will be written as
$(q_0,{\bf q})$. The scalar product in $\Ro^3$ is written as
$\displaystyle
{\bf k}\cdot {\bf q}
=\sum_{\alpha=1}^3{\bf k}_\alpha {\bf q}_\alpha.$
We use the abbreviation $\displaystyle
\partial^\mu\equiv{\partial\ \over\partial q_\mu}.$
The (pseudo)-scalar product of two elements $\psi$ and $\phi$
of a (pseudo)-Hilbert space is denoted $\langle\phi|\psi\rangle$,
linear in $\psi$ and anti-linear in $\phi$.
Throughout the paper, operators carry a hat. The
conjugate of $\hat A$ is denoted $\hat A^*$.

\section*{Structure of the paper}

The next section deals with classical electromagnetism. The vector
potential $A_\mu(q)$ is represented by Fourier coefficients
$a_\mu({\bf k})$. The classical wave functions $\psi_\mu({\bf k})$
are introduced and the duality between classical wave functions
$\psi_\mu({\bf k})$ and Fourier coefficients $a_\mu({\bf k})$ is
established. Section \ref{sectqd} describes the free photon field
as a covariance system. Correlation functions determining the
vacuum state are given explicitly. The field operators $\hat
A(\psi)$ and $\hat F(a)$ live in the corresponding
G.N.S.-representation. Section \ref{sectfr} discusses the standard
Fock representation of the free photon field. Section \ref{sectpi}
starts from Poincar\'e invariance to derive properties of the free
photon. In the final section conclusions are drawn.

\section {Classical electromagnetism}
\label{sectem}

The whole section deals with the classical radiation field. The
vector potential $A_\mu(q)$ is replaced by Fourier coefficients
$a_\mu({\bf k})$. The test functions $f_\mu(q)$, used to smear out the
vector potential $A_\mu(q)$, are replaced by the classical
wave functions $\psi_\mu({\bf k})$. Finally, a duality between
$a_\mu({\bf k})$ and $\psi_\mu({\bf k})$ is established.

\section* {Smeared-out fields}

The classical electromagnetic field is described by the vector
potential $A(q)$. It has four components $A_\mu(q)$, $\mu=0,1,2,3$,
each of which is a function of position $q$ in $\Ro^4$. We assume
that the Lorenz gauge
\be
\partial^\mu A_\mu(q)=0
\label{Lorenzgauge}
\ee
is satisfied. Then the Maxwell equations for the free electromagnetic
field can be written as a set of four equations
\be
\partial^\nu \partial_\nu A_\mu(q)=0.
\label {KGeq}
\ee
It is necessary to smear out $A$ using
test functions. Given real-valued test functions $f_\mu(q)$,
let
\be
f(A)=\int_{\Ro^4}\hbox{ d}q\,f^\mu(q)A_\mu(q).
\label{smearout}
\ee
The functions $f(A)$ will become observables of the free photon
field.

The electric field ${\bf E}$ and the magnetic field ${\bf B}$ are
related to the vector potential $A$ by
\bea {\bf E}_\alpha(q)
&=&-\partial^\alpha A_0(q)
-\partial^0 A_\alpha,\cr
\qquad {\bf
B}_\alpha(q)
&=&\sum_{\beta,\gamma=1}^3\varepsilon_{\alpha\beta\gamma}
\partial^\beta A_\gamma(q) \label{4.1}
\eea
($\varepsilon_{\alpha\beta\gamma}$ is the fundamental antisymmetric
tensor).

Note that the smeared-out electromagnetic fields can be obtained
from the smeared-out vector potential. Indeed one has
\bea
\int_{\Ro^4}\hbox{ d}q\,\sum_{\alpha=1}^3f_\alpha(q){\bf
E}_\alpha(q) &=&-\int_{\Ro^4}\hbox{ d}q\,
\sum_{\alpha=1}^3f_\alpha(q) \partial^\alpha A_0(q)\cr
& & -\int_{\Ro^4}\hbox{ d}q\,
\sum_{\alpha=1}^3f_\alpha(q) \partial^0 A_\alpha(q)\cr
&=&\int_{\Ro^4}\hbox{ d}q\,
A_0(q)\sum_{\alpha=1}^3 \partial^\alpha f_\alpha(q)\cr
& & +\int_{\Ro^4}\hbox{ d}q\,
\sum_{\alpha=1}^3 A_\alpha(q)\partial^0 f_\alpha(q)\cr
&=&g(A)
\eea
with
\bea
g_0(q)&=&\sum_{\alpha=1}^3 \partial^\alpha f_\alpha(q) \quad\hbox{ and }\quad
g_\alpha(q)=-\partial^0 f_\alpha(q),
\eea
and, similarly,
\bea \int_{\Ro^4}\hbox{
d}q\,\sum_{\alpha=1}^3f_\alpha(q){\bf B}_\alpha(q)
&=&\int_{\Ro^4}\hbox{ d}q\,
\sum_{\alpha=1}^3f_\alpha(q)\sum_{\beta,\gamma=1}^3\varepsilon_{\alpha\beta\gamma}
\partial^\beta A_\gamma(q)\cr
&=&-\int_{\Ro^4}\hbox{ d}q\,
\sum_{\alpha,\beta,\gamma=1}^3\varepsilon_{\alpha\beta\gamma}
\left(\partial^\beta f_\alpha(q)\right) A_\gamma(q)\cr
&=&h(A)
\eea
with 
\bea h_0(q)&=&0 \quad\hbox{ and }\quad
h_\gamma(q)=\sum_{\alpha,\beta=1}^3\varepsilon_{\alpha\beta\gamma}
\partial^\beta f_\alpha(q).
\eea

\section*{Fourier coefficients}

Equation (\ref{KGeq}) can be solved by Fourier transformation. Let
\be
\tilde A_\mu(k)=
(2\pi)^{-2}\int_{\Ro^4}\hbox{ d}q
\,\exp\left(ik^\nu q_\nu\right) A_\mu(q).
\label{(1.5)}
\ee
Then (\ref{KGeq}) becomes
$k^\nu k_\nu\tilde A_\mu(k)=0$.
Hence $\tilde A_\mu(k)$ differs from zero only if
$k^\nu k_\nu=0$.
Therefore $A_\mu$ is of the form
\bea
A_\mu(q)&=&(2\pi)^{-2}\int_{\Ro^4}\hbox{ d}k\,
\exp\left(-ik^\nu q_\nu\right)
\tilde A_\mu(k)\delta(k^\sigma k_\sigma)\cr
&=&(2\pi)^{-2}\int_{\Ro^4}\hbox{ d}k\,
\exp\left(-ik^\nu q_\nu\right)
\tilde A_\mu(k)\frac{1}{2|{\bf k}|}\,
\left(\delta(k^0-|{\bf k}|)+\delta(k^0+|{\bf k}|)\right)\cr
&=&(2\pi)^{-2}\int_{\Ro^3}\hbox{ d}{\bf k}\,\frac{1}{2|{\bf k}|}\,
\exp(i{\bf k}\cdot {\bf q})\cr
& &\times
\left(e^{-i|{\bf k}|q_0}\tilde A_\mu(|{\bf k}|, {\bf k})
+e^{i|{\bf k}|q_0}\tilde A_\mu(-|{\bf k}|, {\bf k})
\right).\cr
& &
\eea
Here we use the notation
$|{\bf k}|=\sqrt{\sum_{\alpha=1}^3k_\alpha^2}$.
We obtain
\be
A_\mu(q)
=(2\pi)^{-3/2}\int_{\Ro^3}\hbox{ d}{\bf k}\,\frac{1}{2|{\bf k}|}\,
e^{i{\bf q}\cdot {\bf k}}
\bigg[
e^{-iq_0|{\bf k}|}\avar_\mu({\bf k})
+e^{iq_0|{\bf k}|}\overline {\avar_\mu(-{\bf k})}
\bigg]
\label{fcdef}
\ee
with $\avar_\mu({\bf k})=(2\pi)^{-1/2}\tilde A_\mu(|{\bf k}|, {\bf k})$.
Note that automatically
any vector potential of the form (\ref{fcdef}) satisfies the
wave equations (\ref{KGeq}).
Expression (\ref{smearout}), in combination with (\ref{fcdef}), becomes
\be
f(A)=\sqrt{2\pi}
\int_{\Ro^3}\hbox{ d}{\bf k}\,\frac{1}{2|{\bf k}|}\,
\bigg[
\avar^\mu({\bf k})
\overline{\tilde f_\mu(|{\bf k}|, {\bf k})}
+\overline{\avar^\mu({\bf k})}
\tilde f_\mu(|{\bf k}|, {\bf k})
\bigg]
\label{(1.8)}
\ee
with
\be
\tilde f_\mu(k)=(2\pi)^{-2}
\int_{\Ro^4}\hbox{ d}q\,
\exp(ik^\nu q_\nu)f_\mu(q).
\label{(1.9)}
\ee
Note that only the values of $\tilde f_\mu$ on the light cone
$\{k\in\Ro^4:\ k^\mu k_\mu=0\}$ are of importance.

From the Lorenz condition (\ref{Lorenzgauge}) follows
\be |{\bf
k}|\avar_0({\bf k}) =\sum_{\alpha=1}^3{\bf k}_\alpha
\avar_\alpha({\bf k}). \label{(1.14)}
\ee
This expression can be
used to calculate $\avar_0({\bf k})$ in function of
$\avar_\alpha({\bf k})$.

\section* {Classical wave functions}

The first gauge problem that arises is that two different sets of
test functions $f_\mu(q)$ and $g_\mu(q)$ may define functions
$f(A)$ and $g(A)$ which coincide on all vector potentials $A$ that
satisfy (\ref{Lorenzgauge}) and (\ref{KGeq}). To avoid this
non-uniqueness we make use of the so-called classical wave
functions of the photon. Given test functions $f_\mu$ the
classical wave functions $\psi_\mu$ are defined by
\bea
\psi_\mu({\bf k}) &=&\sqrt{2\pi}\tilde f_\mu(|{\bf k}|,
{\bf k})\cr
&=&(2\pi)^{-3/2}\int_{\Ro^4}\hbox{ d}q\,
\exp\left(iq_0|{\bf k}|-i{\bf q}\cdot {\bf k}\right) f_\mu(q).
\label{classwavefun}
\eea
Note that these are complex functions over $\Ro^3$. Two sets
of test functions $f_\mu$ and $g_\mu$ can give rise to the same
classical wave functions $\psi_\mu$. In fact, this will be the
case if and only if $f(A)=g(A)$ for all $A$ satisfying
(\ref{Lorenzgauge}) and (\ref{KGeq}). Indeed, (\ref{(1.8)}) can
be written as
\bea f(A) &=&\int_{\Ro^3}\hbox{ d}{\bf
k}\,\frac{1}{2|{\bf k}|}\, \bigg[ \avar^\mu({\bf k})
\overline{\psi_{\mu}({\bf k})} +\overline{\avar^\mu({\bf k})}
\psi_{\mu}({\bf k}) \bigg]=-2\Re\langle \avar|\psi\rangle
\label{Afexpr} 
\eea
where the bilinear form
$\langle\cdot|\cdot\rangle$ is given by
\be \langle
\avar|\psi\rangle =-\int_{\Ro^3}\hbox{ d}{\bf k}\,
\frac{1}{2|{\bf k}|}\, \overline {\avar^\mu({\bf k})}
\psi_\mu({\bf k}). \label{bilin}
\ee
This shows that $f(A)$
depends only on the classical wave functions $\psi$ and on the
Fourier coefficients $\avar$.

\section*{Duality}

The Lorenz gauge (\ref{Lorenzgauge}) does not suffice to fix uniquely
the vector potential $A$ corresponding with a given electromagnetic field.
The gauge transformations $A_\mu\rightarrow A'_\mu$ with
\be
A'_\mu=A_\mu+\partial_\mu\chi
\label{gauge}
\ee
with $\chi(q)$ an arbitrary solution of $\partial^\mu\partial_\mu\chi=0$
leave the electric and magnetic fields invariant. We use this
gauge freedom to derive a condition on the classical wave functions.
If two fields $A$ and $A'$ differ only
by a gauge transformation as given by (\ref{gauge}), then no classical
wave function $\psi$ should be able to distinguish $A$ from $A'$,
i.e.~$\Re\langle \avar|\psi\rangle=\Re\langle \avar'|\psi\rangle$
should hold. The result of this condition, deduced below, is
\be
|{\bf k}|\psi_0({\bf k})
=\sum_{\alpha=1}^3{\bf k}_\alpha\psi_\alpha({\bf k}).
\label{fermigauge}
\ee
This condition implies a duality between
Fourier coefficients $\avar({\bf k})$ determining the vector potential $A$
and classical wave functions $\psi({\bf k})$ determining test functions $f$.
Indeed, both are sets of four
complex functions satisfying similar conditions (\ref{(1.14)})
respectively (\ref{fermigauge}).

Because $\chi$ is a solution of $\partial^\mu\partial_\mu\chi=0$
it can be written as (see (\ref{fcdef}))
\be
\chi(q)=\int_{\Ro^3}\hbox{ d}{\bf k}\, \frac{1}{2|{\bf k}|}\,
e^{i{\bf q}\cdot {\bf k}} \bigg[ e^{-iq_0|{\bf k}|}c({\bf k})
+e^{iq_0|{\bf k}|}\overline{c(-{\bf k})} \bigg],
\ee
with $c({\bf
k})$ an arbitrary complex function of ${\bf k}\in\Ro^3$. From
(\ref{gauge}) and (\ref{fcdef}) follows then that
\bea
\avar'_\alpha({\bf k})&=&\avar_\alpha({\bf k})+ic({\bf k}){\bf
k}_\alpha, \qquad\alpha=1,2,3.
\eea
Using (\ref{bilin}) and
(\ref{(1.14)}) one obtains
\bea \langle\avar|\psi\rangle
&=&-\int_{\Ro^3}\hbox{ d}{\bf k}\,\frac{1}{2|{\bf k}|}\,
\left(\overline {\avar_0({\bf k})}\psi_0({\bf k})
-\sum_{\alpha=1}^3\overline {\avar_\alpha({\bf k})}
\psi_\alpha({\bf k}) \right)\cr
&=&-\int_{\Ro^3}\hbox{ d}{\bf k}\, \frac{1}{2|{\bf k}|^2}\,
\sum_{\alpha=1}^3\overline{a_\alpha({\bf k})}
\left(\psi_0({\bf k}) {\bf k}_\alpha -|{\bf k}|\psi_\alpha({\bf k})\right)
\eea
and a similar expression for
$\langle\avar'|\psi\rangle$. Hence the condition
$\Re\langle\avar'|\psi\rangle= \Re\langle\avar|\psi\rangle$ yields
\be
0=\Re\int_{\Ro^3}\hbox{ d}{\bf k}\, \frac{1}{2|{\bf k}|^2}\,
\sum_{\alpha=1}^3i{\bf k}_\alpha \overline{c({\bf k})}
\left(\psi_0({\bf k}) {\bf k}_\alpha -|{\bf k}|\psi_\alpha({\bf
k})\right).
\ee
Because the latter should hold for all choices of
$c({\bf k})$ one concludes that (\ref{fermigauge}) holds.

\section*{Radiation gauge}

Note that the scalar product $\langle a|\psi\rangle$ is
degenerate. Indeed, condition (\ref{fermigauge}), to be satisfied
by classical wave functions, was derived precisely by requiring
that, if a gauge transformation maps $a$ onto $b$ then $\langle
a|\psi\rangle=\langle b|\psi\rangle$ holds for all $\psi$. By
duality, we say that $\psi$ and $\phi$ are equivalent if $\langle
a|\psi\rangle=\langle a|\phi\rangle$ holds for all $a$. As noted
by Carey et al\cite{CGH77}, in each set of equivalent $\psi$ one
can select a unique representative $\psi$ satisfying
\be
\psi_0({\bf k})=0 \qquad\hbox{ and }\quad \sum_{\alpha=1}^3{\bf
k}_\alpha\psi_\alpha({\bf k})=0. \label{radgauge}
\ee
See Appendix
C. These conditions are called the {\sl radiation gauge}. However,
by selecting such a representative one breaks the property of
manifest Lorentz invariance. Therefore, we will use this gauge
only to discuss the physical content of certain formulas.

\section{Quantum description}
\label{sectqd}

This section gives a description of the electromagnetic field as
a quantum system. We start from explicit correlation functions
and use the generalized GNS-theorem to construct a representation
in Hilbert space.

\section*{Covariance approach}

In the previous section the smeared-out vector potential $f(A)$
could be written as $-2\Re\langle \avar|\psi\rangle$ (see
(\ref{Afexpr})), where $\avar({\bf k})$ are Fourier coefficients
representing the vector potential $A_\mu(q)$ and $\psi({\bf k})$
is a classical wave function representing the test functions
$f_\mu(q)$. In the quantum theory both $a({\bf k})$ and $\psi({\bf
k})$ become operators in Hilbert space. They will be denoted $\hat
F(a)$ and $\hat A(\psi)$, respectively.

We want to derive the quantum description of the electromagnetic
field in a way similar to the quantum description of a
single particle. The quantity corresponding with a function $f(q)$
of the position $q$ of the particle is the function $f(A)$
considered as a function of the vector potential $A_\mu$.
Hence, in the obvious quantum description quantum mechanical
wave functions would be complex square integrable
functions of $A_\mu$ (replacing $q$-dependent functions)
and the $f(A)$ is mapped onto an operator $f(\hat A)$
(replacing $f(\hat q)$, with $\hat q\psi(q)=q\psi(q)$).
A more common notation, replacing $f(\hat A)$, is $\hat A(\psi)$,
with $\psi$ the classical wave function corresponding with $f$.
As discussed in the introduction, the problem with this approach
is that the operators $\hat A(\psi)$ are expected not to
be mutually commuting, so that they cannot be simple multiplication
operators, as in the case of quantum mechanics of a single particle.
The solution adopted here is to see the operators $\hat A(\psi)$
as generators of the group of adding test functions. An
additional advantage of this point of view is that it is then
natural to consider also the group of adding fields. The generators
of the latter group are the operators $\hat F(a)$.
Another advantage is that the resulting formalism is very close to
the $C^*$-algebraic approach using Weyl algebras\cite{CGH77,PD90}.

\section*{Correlation functions}

The classical wave functions $\psi$ form a linear space, denoted
$H$. The Fourier coefficients $\avar$ belong to the dual space
$H^*$. In what follows we will consider $H^*$ as a real linear
space, and not a complex one, because only multiplication of
$a_\mu$ with a real constant corresponds with multiplication of
the vector potential $A_\mu$ with the same constant. As a
consequence, also $H$ will be considered as a real linear space.

Consider $H^*\times H$ as an additive group. A state of the
covariance system $(\Co,H^*\times H,\Io)$ is determined by
correlation functions ${\cal F}(a,\psi;b,\phi)$. We make the
following choice:
\bea
{\cal F}(a,\psi;b,\phi)&=&
\exp\left(-\frac{i}{2\eta}\Im\langle b+i\eta\phi|a+i\eta\psi\rangle
\right)\cr
&\times& \exp\left(-\frac{1}{4\eta}\,\langle b-a+i\eta(\phi-\psi)|
b-a+i\eta(\phi-\psi)\rangle\right)
\label{corfun}
\eea
with $\eta$ a positive number.
The proof that
these correlation functions have the necessary properties to
define a state of $(\Co,H^*\times H,\Io)$ is given in Appendix D.
The generalized GNS-theorem\cite {NK00} implies that there exists
a projective representation $\hat W(a,\psi)$ of $H^*\times H$ in a
Hilbert space $\cal H$, and a normalized wave function $\Omega$ in
$\cal H$, for which
\be {\cal F}(a,\psi;b,\phi) =\langle \hat
W(b,\phi)^*\Omega|\hat W(a,\psi)^*\Omega\rangle \label{repfun}
\ee
holds.

\section*{The cocycle}

From the {\sl ansatz}
\be \hat W(a,\psi)\hat W(b,\phi)
=\exp\left(\frac{i}{2}\,s(a,\psi;b,\phi)\right) \hat
W(a+b,\psi+\phi) \label{weyl}
\ee
follows, using that $\hat
W(a,\psi)^*=\hat W(-a,-\psi)$,
\bea {\cal F}(a,\psi;b,\phi)
&=&\langle \Omega|\hat W(b,\phi)\hat W(-a,-\psi)\Omega\rangle\cr
&=&\exp\left(-\frac{i}{2}\,s(b,\phi;a,\psi)\right) \langle
\Omega|\hat W(b-a,\phi-\psi)\Omega\rangle\cr
&=&\exp\left(-\frac{i}{2}\,s(b,\phi;a,\psi)\right) {\cal
F}(a-b,\psi-\phi;0,0).
\eea
Using the definition of $\cal F$ one
obtains
\be s(a,\psi;b,\phi)=\eta^{-1}\Im\langle a+i\eta\psi|b+i\eta\phi\rangle.
\ee
This function $s(a,\psi;b,\phi)$ is a symplectic form, as it
should be. It is anti-symmetric under exchange of $(a,\psi)$ and
$(b,\phi)$. It is real linear in its arguments. Note that it is
degenerate because $\langle\cdot|\cdot\rangle$ is degenerate. The
function $\exp\left((i/2)s(a,\psi;b,\phi)\right)$, appearing in
(\ref{weyl}), is a cocycle of the additive group $H^*\times H$.

\section*{Field operators}

The operators $\hat A(\psi)$ and $\hat F(a)$ are introduced as
self-adjoint operators satisfying 
\bea \hat
W(0,\lambda\psi)&=&\exp\left(i\lambda\hat A(\psi)\right)
\quad\hbox{ and }\quad \hat W(\lambda a,0)=\exp\left(i\lambda \hat
F(a)\right) \label{opdef}
\eea
for all real $\lambda$. The
commutation relations for these operators can be obtained from
\be
\hat W(a,\psi)\hat W(b,\phi)=\exp\left(is(a,\psi;b,\phi)\right)
\hat W(b,\phi)\hat W(a,\psi) \label{weylccr}
\ee
by inserting real
numbers, as in (\ref{opdef}), and taking derivatives. One obtains
\bea
\left[\hat F(a),\hat F(b)\right]
&=&-i\eta^{-1}\Im\langle a|b\rangle\cr
\left[\hat F(a),\hat A(\phi)\right]
&=&-i\Re\langle a|\phi\rangle\cr
\left[\hat A(\psi),\hat A(\phi)\right]
&=&-i\eta\Im\langle \psi|\phi\rangle.
\label{faccr}
\eea
Comparison with the traditional result (\ref{emfcr})
gives $\eta=2$.

The unitary operator $\hat W(a)$ implements adding (or subtracting) a
field. Indeed, one calculates
\bea \hat W(a,0)\hat A(\psi)\hat
W(-a,0) &=&-i\frac{{\rm d}\,}{{\rm
d}\lambda}\bigg\vert_{\lambda=0} \hat W(a,0)\hat
W(0,\lambda\psi)\hat W(-a,0)\cr
&=&-i\frac{{\rm d}\,}{{\rm
d}\lambda}\bigg\vert_{\lambda=0} \exp\left(i\lambda
s(a,0;0,\psi)\right)\hat W(0,\lambda\psi)\cr
&=&\hat
A(\psi)+\Re\langle a|\psi\rangle\cr
&=&\hat A(\psi)-f(A)
\eea
with
$f_\mu(q)$ the test functions corresponding with $\psi$ and
$A_\mu(q)$ the vector potential corresponding with $a$. Formally,
this can be rewritten as
\be \hat W(a,0)\hat A_\mu(q)\hat
W(-a,0)=\hat A_\mu(q)-A_\mu(q).
\ee
Similarly, the unitary
operator $\hat W(0,\psi)$ implements adding wave functions.
Indeed, one finds 
\bea \hat W(0,\psi)\hat F(a)\hat W(0,-\psi)
&=&\hat F(a)-\Re\langle\psi|a\rangle\cr
&=&\hat F(a)+f(A).
\eea

\section*{Real additivity and identification}

Let us calculate
\begin{eqnarray}
& &\langle\hat W(b,\phi)^*\Omega|\,
\hat A(\chi)\hat W(a,\psi)^*\Omega\rangle\cr
&=&i\frac{\partial\,}{\partial \lambda}\bigg|_{\lambda=0}
\langle\hat W(b,\phi)^*\Omega|\,
\hat W(0,\lambda\chi)^*\hat W(a,\psi)^*\Omega\rangle\cr
&=&i\frac{\partial\,}{\partial \lambda}\bigg|_{\lambda=0}
\exp\left(-(i\lambda/2)\Re\langle\chi|\,a+i\eta\psi\rangle\right)
{\cal F}(a,\psi+\lambda\chi;b,\phi)\cr
&=&\frac{1}{2}\left[
\langle a+i\eta\psi|\,\chi\rangle+\langle\chi|\,b+i\eta\phi\rangle
\right]
\langle\hat W(b,\phi)^*\Omega|\,\hat W(a,\psi)^*\Omega\rangle
\label{amatrix}
\end{eqnarray}
and, similarly,
\begin{eqnarray}
& &\langle\hat W(b,\phi)^*\Omega|\,
\hat F(d)\hat W(a,\psi)^*\Omega\rangle\cr
&=&\frac{i}{2\eta}\left[
\langle d|\,b+i\eta\phi\rangle -\langle a+i\eta\psi|\, d\rangle
\right]
\langle\hat W(b,\phi)^*\Omega|\,\hat W(a,\psi)^*\Omega\rangle.
\end{eqnarray}

These expressions show that the operators $\hat A(\chi)$ and 
$\hat F(d)$ are real linear functions. Moreover, comparison of 
the two expressions yields $\eta\hat F(i\chi)=\hat A(\chi)$ for 
all $\chi$. One concludes that in the Hilbert space 
representation determined by the correlation functions 
(\ref{corfun}) the generators of adding fields, respectively of 
adding wave functions, coincide.

\section{Fock representation}
\label{sectfr}

In this section the Hilbert space representation
determined by the correlation functions (\ref{corfun})
is identified with the Fock space in which photon states
are created by repeated application of creation operators
onto the vacuum state.

\section*{Creation and annihilation operators}

From (\ref{amatrix}) follows that
\be \langle \hat W(b,\phi)^*\Omega|\hat A(\psi)\Omega\rangle
=\frac{1}{2}\langle \hat W(b,\phi)^*\Omega|\Omega\rangle \langle
\psi|b+i\eta\phi\rangle.
\label{aid}
\ee
Hence one has
\bea \langle
\hat W(b,\phi)^*\Omega|(\hat A(\psi)-i\hat A(i\psi))\Omega\rangle
&=&0.
\eea
Since $b$ and $\phi$ are arbitrary this implies that
\bea (\hat A(\psi)-i\hat A(i\psi))\Omega=0.
\eea
It is therefore
obvious to define {\sl annihilation operators} $\hat A_-(\psi)$ by
\be
\hat A_-(\psi)=\frac{1}{2}\,\hat A(\psi)-\frac{i}{2}\,\hat
A(i\psi),
\ee
which means that also
\be
\hat
A_-(\psi)=\frac{1}{2}\,\hat A(\psi)+\frac{i\eta}{2}\,\hat F(\psi).
\ee
The latter expression resembles the definition $Q+iP$ of the
annihilation operator by means of a pair of position and momentum
operators, in the context of the harmonic oscillator. Note that
$\hat A_-(\psi)$ is a complex linear function of $\psi$. As shown
above, the annihilation operators satisfy
\be
\hat A_-(\psi)\Omega=0.
\ee
One verifies that these operators are
commuting
\bea
\left[\hat A_-(\psi),\hat A_-(\phi)\right] &=&0.
\eea
The conjugate operator $\hat A_+(\psi)=\hat A_-(\psi)^*$ is
the {\sl creation operator}. From the definition follows
immediately that
\be
\hat A(\psi)=\hat A_+(\psi)+\hat A_-(\psi).
\ee The commutation relations between creation and annihilation
operators are found to be
\bea \left[\hat A_+(\psi),\hat
A_-(\phi)\right] &=&\frac{1}{4}\,\left[\hat A(\psi)+i\hat
A(i\psi), \hat A(\phi)-i\hat A(i\phi)\right]\cr
&=&-\frac{\eta}{2}\langle\psi|\phi\rangle.
\eea
With $\eta=2$ this relation gives to the operator
$\hat A_+(\psi)\hat A_-(\phi)$ the usual interpretation
of number operator.

\section*{One-photon states}

A one-photon state is determined by an element of the Hilbert
space of the form $\hat A_+(\xi)\Omega$, where $\xi$ is a
classical wave function, not equivalent to zero. It is
straightforward to verify that two equivalent classical
wave functions $\psi$ and $\phi$ determine the same one-photon
state. Indeed, by definition they satisfy $\langle
a|\psi\rangle=\langle a|\phi\rangle$ for all $a$. From
(\ref{aid}) then follows that $\hat A(\psi)\Omega=
\hat A(\phi)\Omega$ so that
$\hat A_+(\psi)\Omega=\hat A_+(\phi)\Omega$.
Hence $\phi$ and $\psi$
determine the same wave function in the Hilbert space, and hence,
the same physical state.

A short calculation gives
\bea ||\hat A_+(\psi)\Omega||^2
&=&\langle\Omega|\hat A_-(\psi)\hat A_+(\psi)\Omega\rangle\cr
&=&-\langle\Omega| \left[\hat A_+(\psi),\hat
A_-(\psi)\right]\Omega\rangle\cr
&=&\frac{\eta}{2}\langle\psi|\psi\rangle\cr
&=&\frac{\eta}{2}\int_{\Ro^3}\hbox{ d}{\bf k}\,
\frac{1}{2|{\bf k}|^3}\,
\sum_{\alpha,\beta=1}^3\overline{\psi_\alpha({\bf k})} \left(|{\bf
k}|^2\delta_{\alpha\beta}- {\bf k}_\alpha {\bf k}_\beta\right)
\psi_\beta({\bf k})\cr
&\ge&0. \label{onephnorm}
\eea
The last
steps of this calculation use results of Appendix D. In
particular, $||\hat A_+(\psi)\Omega||=0$ holds if $\psi_\mu({\bf
k})$ is of the form
\bea
\psi_\alpha({\bf k}) =\frac{{\bf
k}_\alpha}{|{\bf k}|}\,\psi_0({\bf k}), \qquad\alpha=1,2,3.
\label{normzero}
\eea
The usual interpretation of this result is
that there do not exist photon states for which the electric and
magnetic fields are not perpendicular to the wave vector $\bf k$.

\section{Poincar\'e invariance}
\label{sectpi}

In this section we study the action of the proper Poincar\'e group
in Fock space. The generators of this group determine physical
quantities like energy, momentum, mass, and spin of the photon.

\section*{Shifts in spacetime}

A shift with vector $x$ in spacetime maps the vector potentials
$A_\mu(q)$ onto vector potentials $A^x_\mu(q)$ given by
\be
A^x_\mu(q)=A_\mu(q-x).
\ee
Using (\ref{fcdef}), one finds that the Fourier
coefficients $a_\mu$ transform into $a^x_\mu$
given by
\be
a^x_\mu({\bf k})=
\exp\left(i|{\bf k}|x_0-i{\bf k}\cdot{\bf x}\right)
a_\mu({\bf k}).
\label{ashift}
\ee
The corresponding transformation of the classical wave functions
$\psi_\mu({\bf k})$ is
\be
\psi^x_\mu({\bf k})=
\exp\left(i|{\bf k}|x_0-i{\bf k}\cdot {\bf x}\right)
\psi_\mu({\bf k}).
\label{phshift}
\ee
With this choice of action
the correlation functions ${\cal F}(a,\psi;b,\phi)$
are invariant under shifts.
This is what we want because the corresponding state
of the system is the vacuum state.
Note that the smeared-out fields $f(A)$ are
invariant under shifts.

A unitary representation of the group of shifts $\Ro^4,+$
is defined by
\be
\hat U(x)\hat W(a,\psi)\Omega=\hat W(a^x,\psi^x)\Omega
\label{unitshiftdef}
\ee
(see Appendix E).
The generators of this representation are denoted $\hat K_\mu$
and are defined by
\be
\hat U(x)=\exp(-ix^\mu \hat K_\mu).
\ee
By convention, the momentum operators $P_\mu$ equal
$\hbar \hat K_\mu$. The energy operator is $cP_0=c\hbar \hat K_0$.
The vector $\Omega$ corresponds with the vacuum state
and is invariant under shifts. In particular,
$\hat K_\mu\Omega=0$ holds. Hence the energy and momentum
of the vacuum are zero, contrary to what is claimed
in textbooks, based on the harmonic oscillator picture
of the photon.

\section*{Energy and momentum of a one-photon state}

Let us calculate
\bea
& &\langle \hat W(b,\phi)^*\Omega|\hat
K_0\hat W(a,\psi)^*\Omega\rangle\cr
&=&i\frac{\partial\,}{\partial
x_0}\bigg\vert_{x=0} \langle \hat W(b,\phi)^*\Omega|\hat U(x)\hat
W(a,\psi)^*\Omega\rangle\cr
&=&i\frac{\partial\,}{\partial x_0}\bigg\vert_{x=0}
{\cal F}(a^x,\psi^x;b,\phi)\cr
&=&-{\cal F}(a,\psi;b,\phi)
i\frac{\partial\,}{\partial x_0}\bigg\vert_{x=0}
\bigg(\frac{i}{2\eta}\Im\langle b+i\phi|a^x+i\psi^x\rangle\cr
& &
+\frac{1}{4\eta}\langle b-a^x+i\eta(\phi-\psi^x)|
b-a^x+i\eta(\phi-\psi^x)\rangle\bigg)\cr
&=&\frac{1}{2\eta}{\cal
F}(a,\psi;b,\phi) \langle a+i\eta\psi|\,|{\bf k}|\,(b+i\eta\phi)\rangle.
\label{explknot}
\eea
Similarly, one shows that, with
$\alpha=1,2,3$,
\be
\langle \hat W(b,\phi)^*\Omega|\hat
K_\alpha\hat W(a,\psi)^*\Omega\rangle =\frac{1}{2\eta}{\cal
F}(a,\psi;b,\phi) \langle a+i\eta\psi| {\bf k}_\alpha(b+i\eta\phi)\rangle.
\label{explkalpha}
\ee
Consider now a one-photon state. From (\ref{explknot}) and
(\ref{explkalpha}) it follows that
\be
K_0\hat
A_+(\psi)\Omega=\hat A_+(|{\bf k}|\psi)\Omega \quad\hbox{ and
}\quad K_\alpha\hat A_+(\psi)\Omega=\hat A_+({\bf
k}_\alpha\psi)\Omega.
\ee
To see this, use that
\begin{equation}
\langle W(b,\phi)^*|\,A_+(\psi)\Omega\rangle
=\frac{1}{2}\langle\psi|b+i\eta\phi\rangle
{\cal F}(0,0;b,\phi).
\end{equation}

Here we work with photon states smeared
out with classical wave functions. It is tradition to associate
the notion of photon with states that are not smeared out. These
are idealized states which are not represented by wave functions
in Hilbert space. In order to approach such a photon state we have
to select classical wave functions that converge to a Dirac
measure concentrated at a single wave vector ${\bf k}$. The energy
of such a photon is then equal to $\hbar c|{\bf k}|$, the momentum
is equal to $\hbar {\bf k}$. In particular, this implies that the
mass of such an idealized photon is exactly equal to zero.

\section*{Lorentz transformations}

The discussion of Lorentz transformations is not very easy
because both Fourier coefficients $a_\mu({\bf k})$ and classical
wave functions $\psi_\mu({\bf k})$ depend on a wave vector ${\bf
k}$ in $\Ro^3$, instead of covariant vectors in $\Ro^4$,
and do not obey easy transformation rules.
However, both the vector potential $A_\mu(q)$ and the
test functions $f_\mu(q)$ transform as vectors so that the
smeared-out vector potential $f(A)$ is invariant under Lorentz
transformations. Since $f(A)=-\Re\langle a|\psi\rangle$ holds,
this shows that the pseudo-scalar product is invariant under
Lorentz transformations. Hence the correlation functions
(\ref{corfun}) are invariant under Lorentz transformations.

Let $\Lambda$ denote a proper Lorentz transformation.
Under its action the vector potential $A_\mu(q)$ transforms
into $A'_\mu(q)$ given by
\be
A'_\mu(q)=\Lambda_\mu^{\,\nu}A_\nu(\Lambda^{-1} q).
\ee
Assume first that $\Lambda$ is a spatial rotation described
by the 3-by-3 matrix $R$. Then the Fourier coefficients
$\avar_\mu({\bf k})$ transform into $\avar'_\mu({\bf k})$
given by
\be
\avar'_\mu({\bf k})=\Lambda_\mu^{\,\nu}\avar_\nu(R{\bf k}).
\ee
Next consider a boost in direction 3.
The non-zero matrix elements are
$\Lambda_{00}=\Lambda_{33}=\cosh(\chi)$,
$\Lambda_{03}=\Lambda_{30}=\sinh(\chi)$,
$\Lambda_{11}=\Lambda_{22}=1$.
Then one obtains
\bea
\avar'_0({\bf k})&=&\cosh(\chi)\avar_0({\bf k}')
-\sinh(\chi)\avar_3({\bf k}')\cr
\avar'_1({\bf k})&=&\avar_1({\bf k}')\cr
\avar'_2({\bf k})&=&\avar_2({\bf k}')\cr
\avar'_3({\bf k})&=&-\sinh(\chi)\avar_0({\bf k}')
+\cosh(\chi)\avar_3({\bf k}')
\label{pboost}
\eea
with
\be
{\bf k}'=({\bf k}_1,{\bf k}_2,\cosh(\chi){\bf k}_3+\sinh(\chi)|{\bf k}|).
\ee
Together, the spatial rotations and the boosts in direction 3
generate the proper Lorentz group. Hence, the above formulas
represent the action of the proper Lorentz group on the Fourier coefficients
$a({\bf k})$. The classical wave functions $\psi({\bf k})$
transform in a similar way. A unitary operator $\hat V(\Lambda)$
is now defined by
\bea
\hat V(\Lambda)\hat W(a,\psi)\Omega
&=&\hat W(a',\psi')\Omega.
\label{unreplg}
\eea
These unitary operators form a representation
of the proper Lorentz group. To show this
one uses the same arguments as in case of the
group of shifts.

\section*{The generators of spatial rotations}

The six generators of the proper Lorentz group are denoted
$\hat M_{\mu\nu}=-\hat M_{\nu\mu}$. Three of them correspond with
spatial rotations, the other three with boosts.

Consider now a rotation by an angle $\chi$ around the third
coordinate axis. The Fourier coefficients transform like
\bea
\avar'_0({\bf k}) &=&\avar_0({\bf k}')\cr
\avar'_1({\bf k})
&=&\cos(\chi)\avar_1({\bf k}')-\sin(\chi)\avar_2({\bf k}')\cr
\avar'_2({\bf k}) &=&\sin(\chi)\avar_1({\bf
k}')+\cos(\chi)\avar_2({\bf k}')\cr
\avar'_3({\bf k})
&=&\avar_3({\bf k}') \label{pboostrot}
\eea
with
\be
{\bf
k}'=(\cos(\chi){\bf k}_1+\sin(\chi){\bf k}_2,
-\sin(\chi){\bf k}_1+\cos(\chi){\bf k}_2,{\bf k}_3).
\ee
We calculate
\bea
& &
\langle\hat W(b,\phi)^*\Omega|\hat M_{12}\hat W(a,\psi)^*\Omega\rangle\cr
&=&i\frac{\partial\,}{\partial \chi}\bigg\vert_{\chi=0} \langle
\hat W(b,\phi)^*\Omega|V(\Lambda)\hat W(a,\psi)^*\Omega\rangle\cr
&=&i\frac{\partial\,}{\partial \chi}\bigg\vert_{\chi=0}
{\cal F}(a',\psi';b,\phi)\cr
&=&-\frac{i}{2\eta} {\cal F}(a,\psi;b,\phi)
\frac{\partial\,}{\partial \chi}\bigg\vert_{\chi=0}\cr
& &\bigg(i\Im\langle b+i\eta\phi|a'+i\eta\psi'\rangle
+\frac{1}{2}\langle b-a'+i\eta(\phi-\psi')|b-a'+i\eta(\phi-\psi')\rangle
\bigg)\cr
&=&\frac{1}{2\eta}{\cal F}(a,\psi;b,\phi)
\langle a+i\eta\psi|(S_{12}+L_{12})(b+i\eta\phi)\rangle\cr
& &
\label{genrot}
\eea
with $S_{\mu\nu}$ the 4-by-4-matrix with $i$ at
position $\mu,\nu$, $-i$ at position $\nu,\mu$, and
zeroes everywhere else, and with
\be
L_{12}=i
\left({\bf k}_1\frac{\partial\,}{\partial {\bf k}_2}
-{\bf k}_2\frac{\partial\,}{\partial {\bf k}_1}
\right).
\ee

\section*{The generators of boosts}

Now let $\Lambda$ be a boost in direction 3, as given by
(\ref{pboost}). The corresponding generator $M_{03}$ is calculated
as follows.
\bea
& &\langle \hat W(b,\phi)^*\Omega|\hat M_{03}\hat W(a,\psi)^*\Omega\rangle\cr
&=&-i\frac{\partial\,}{\partial\chi}\bigg\vert_{\chi=0}
\langle \hat W(b,\phi)^*\Omega|V(\Lambda)\hat W(a,\psi)^*\Omega\rangle\cr
&=&-i\frac{\partial\,}{\partial\chi}\bigg\vert_{\chi=0}
{\cal F}(a',\psi';b,\phi)\cr
&=&-\frac{i}{2\eta}{\cal F}(a,\psi;b,\phi)
\frac{\partial\,}{\partial\chi}\bigg\vert_{\chi=0}\cr
& &
\bigg(
-i\Im\langle b+i\eta\phi|a'+i\eta\psi'\rangle
-\frac{1}{2}\langle b-a'+i\eta(\phi-\psi')|
b-a'+i\eta(\phi-\psi')\rangle \bigg)\cr
&=&-\frac{1}{2\eta}{\cal F}(a,\psi;b,\phi)
\langle a+i\eta\psi|(S_{03}-L_{03})(b+i\eta\phi)\rangle.
\label{genboost}
\eea
with
\be
L_{0\alpha}=i|{\bf k}|\frac{\partial\,}{\partial {\bf k}_\alpha}.
\ee

\section*{Spin of the photon}

From expressions (\ref{genrot}, \ref{genboost}) it is clear that
there are two different types of contributions to the generators
$M_{\mu\nu}$ of the Lorentz group. These are called the spin part
$\hat S$, respectively the orbital part $\hat L$. The spin
contribution originates from the vector character of the
electromagnetic vector potential $A_\mu(q)$, the orbital part
follows from the transformation of Minkowski space. The operators
$\hat L_{23}$, $\hat L_{31}$ and $\hat L_{12}$ are the components
of angular momentum, the operators $\hat S_{23}$, $\hat S_{31}$
and $\hat S_{12}$ are the components of the spin of the photon.
Note that these notions are not covariant. Worse is that the
splitting of $\hat M$ into $\hat S$ and $\hat L$ is not gauge
invariant. This implies that the components of $\hat S$ and $\hat
L$ are not physically observable. Hence one could say that the
mechanical spin of the photon is not observable. See Jauch and
Rohrlich\cite {JR80} for a discussion of these points.

However, it is common to say that the photon is a spin-1 particle.
In order to understand this statement let us assume that a
one-photon state $\hat A_+(\psi)\Omega$ is an eigenstate of the
operator $\hat S_{12}$
\be
\hat S_{12}\hat A_+(\psi)\Omega=\lambda
\hat A_+(\psi)\Omega.
\ee
From
\be
\langle \hat
W(b,\phi)^*\Omega|\hat S_{12}\hat W(a,\psi)^*\Omega\rangle
=\frac{1}{2\eta}{\cal F}(a,\psi;b,\phi)
\langle a+i\eta\psi|S_{12}(b+i\eta\phi)\rangle
\label{spin12me}
\ee
follows
\be
\langle \hat
W(b,\phi)^*\Omega|\hat S_{12}A_+(\psi)\Omega\rangle
=\frac{1}{2}\langle \hat W(b,\phi)^*\Omega|\Omega\rangle
\langle\psi|S_{12}(b+i\eta\phi)\rangle.
\ee
In combination with
the assumption that the one-photon state is an eigenstate of $\hat
S_{12}$ with eigenvalue $\lambda$ there follows
\be
\lambda\langle \hat W(b,\phi)^*\Omega|A_+(\psi)\Omega\rangle
=\frac{1}{2}\langle \hat W(b,\phi)^*\Omega|\Omega\rangle
\langle\psi|S_{12}(b+i\eta\phi)\rangle.
\ee
On the other hand is
\be
\langle \hat W(b,\phi)^*\Omega|A_+(\psi)\Omega\rangle=\frac{1}{2}
\langle \hat W(b,\phi)^*\Omega|\Omega\rangle
\langle\psi|b+i\eta\phi\rangle.
\ee
Comparison of both expressions
gives the condition
\be
S_{12}\psi({\bf k})=\lambda\psi({\bf k}).
\ee
Now, the eigenvalues of
the matrix $S_{12}$ are +1, -1, and 0 (two-fold degenerated).
Hence, the space of classical wave functions $H$ can be split into
three real-linear subspaces $H_+$, $H_0$, and $H_-$ with the
properties that $S_{12}\psi=\pm\psi$ if $\psi$ is in $H_+$,
respectively in $H_-$, and $S_{12}\psi=0$ if $\psi$ is in $H_0$.
As a consequence, also the one-photon subspace of Fock space can
be written as a direct sum of three subspaces which consist of
eigenvectors of $\hat S_{12}$ corresponding to the eigenvalues
$+1,0,-1$. Spin-operators with a spectrum $+1,0,-1$
are associated with spin-1 particles.

\section*{Polarization of the photon}

As stated earlier in section \ref{sectem}, each class of
equivalent classical wave functions contains a representative $\psi$
satisfying $\psi_0=0$. None of these representatives belongs to
$H_0$. Indeed, if $S_{12}\psi({\bf k})=0$ holds for all
${\bf k}$ then $\psi_1=\psi_2=0$ and $|{\bf k}|\psi_0({\bf k})=
{\bf k}_3\psi_3({\bf k})$. But because of $\psi_0=0$ also
$\psi_3=0$ follows. Hence, if a representative $\psi$ belongs
to $H_0$ then all of its components are zero. Classical wavefunctions
in $H_\pm$ satisfy
\be
0=({\bf k}_1\mp i{\bf k}_2)\psi_1({\bf k})={\bf k}_3\psi_3({\bf k})
\quad\hbox{ and }\psi_2({\bf k})=\mp i\psi_1({\bf k})
\ee
with $\psi_1({\bf k})$ and $\psi_2({\bf k})$ not identically zero.
The only solutions of these conditions
are idealized photons with wave vector
parallel to the third direction and with $\psi_3({\bf k})=0$.
This implies that the electric and magnetic fields lie in the plane
orthogonal to the wave vector $\bf k$. Two
independent solutions are allowed. They correspond with
the two independent polarizations
of the electromagnetic field. A more detailed analysis
can be found in Jauch and Rohrlich\cite {JR80}.

\section{Conclusions}

We have shown in this paper that the standard theory of the free
photon field can be derived within the covariance approach to
quantum mechanics. Typical for this approach is that it starts
from the action of a group in a $C^*$-algebra and from
correlation functions describing a state of the covariance
system. In the case of the free radiation field the $C^*$-
algebra is the algebra of complex numbers, the group is the
group of adding fields times the group of adding test functions.
The Lorenz gauge is used to eliminate part of the redundancy.
The action is trivial. The correlation functions describe a
Lorentz invariant vacuum state. The state vectors of the induced
Fock representation are invariant under the remaining gauge freedoms.

The present approach has several advantages. In the first place
the formalism is mathematically rigorous. The development of
photon theory is crystal clear and there is no need for
hand waving arguments. Both the gauge problem and the problem of
positivity of the scalar product are solved in a satisfactory
manner. The approach is generic. It is obvious how to apply it
to other fields than the electromagnetic one.

We have stressed that there exists a duality between classical
wave functions $\psi$ and Fourier coefficients $a$. As a
consequence of this duality there exist, besides the usual field
operators $\hat A(\psi)$, also operators $\hat F(a)$, labeled with
Fourier transformed vector potentials. However,
in the representation of the electromagnetic vacuum state
the field operators $\hat F(a)$ and $\hat A(\psi)$ coincide.
Hence, this duality has no practical consequences for the
description of the vacuum. We do not know if this degeneracy
continues to exist in other representations of the electromagnetic
radiation field.

The formalism considers photons smeared out with classical wave
functions. These differ from the idealized photons discussed in
most text books. The reason for smearing-out is of course that
the strictly localized objects $\hat A(q)$ cannot be defined as
operators in Fock space, while the smeared-out equivalents $\hat
A(\psi)$ are nicely defined self-adjoint operators.
Finally, the generators of the Lorentz
group can be decomposed into a sum of an orbital part and a spin
part. The space of classical wave functions can be split into
three parts corresponding with spin 1, 0 , and -1 respectively.
Because of gauge freedom only two independent polarizations of
the idealized photons occur.

Up to now, we did not consider electromagnetic fields in
presence of external charges and currents. The first question
that arises in this context is whether all states of the
covariance system of the free radiation field (i.e.~the one used
in the present paper) describe radiation fields, or whether
states can be found which describe fields produced by charges
and currents. If the latter is not the case, then the covariance
system has to be modified. Another topic for further
investigation is the description of massive photons in terms of
the present formalism (see e.g.~section 6-5 of Jauch and
Rohrlich\cite {JR80}). Our ultimate goal is of course a
combination of electron and photon fields within the same
covariance approach.

\section*{Acknowledgement}
We thank Marek Czachor for his interest in the present work.

\section*{Appendix A: Covariance systems}

A covariance system $({\cal A},X,\sigma)$ consists of a
$C^*$-algebra $\cal A$, a locally compact group $X$, and an action
$\sigma$ of this group as automorphisms of $\cal A$. For each
$a\in{\cal A}$ the map $x\in X\rightarrow \sigma_xa$ should be
continuous. In the present paper, the $C^*$-algebra $\cal A$ is
the algebra $\Co$ of complex numbers. In this case the only
possible action of $X$ is the trivial one, leaving the complex
numbers invariant. The resulting covariance system is denoted
$(\Co,X,\Io)$ and is rather trivial. Still, the notions of
state and of representation of a covariance system apply,
and are nontrivial.

A state\cite {NK00} of a covariance system $({\cal A},X,\sigma)$
is determined by correlation functions ${\cal F}(a,x,y)$
depending on $a\in{\cal A}$ and $x,y\in X$. They satisfy
conditions of positivity, normalization, covariance, and continuity.
In the present context, where the $C^*$-algebra is the algebra of
complex numbers, the dependence on elements of $\cal A$ can be
omitted, and the conditions reduce to
\begin{itemize}
\item (positivity)
For all $n>0$ and for all possible choices of
$\lambda_1,\ldots,\lambda_n$ in $\Co$,
of $x_1,\ldots,x_n$ in $X$, is
\be
\sum_{j,k=1}^n\lambda_j\overline{\lambda_k}{\cal F}(x_j,x_k)
\ge 0.
\ee
\item (normalization)
${\cal F}(e,e)=1$ ($e$ is the neutral element of $X$).
\item (covariance)
${\cal F}(xz,yz)={\cal F}(x,y)$ for all $x,y,z$ in $X$.
\item (continuity)
the map $x,y\rightarrow {\cal F}(x,y)$ is
continuous in a neighborhood of the neutral element of $X$.
\end{itemize}

A representation of the covariance system $(\Co,X,\Io)$
is nothing but a projective representation $U$ of the group $X$ as
unitary operators of a Hilbert space $\cal H$, with the property
that the map $x\rightarrow U(x)$ is strongly continuous for $x$
in a neighborhood of the neutral element of $X$. In this context
the generalized G.N.S.-theorem states that for each state of
$(\Co,X,\Io)$, described by the correlation functions
${\cal F}(x,y)$, there exists a representation $U$ of
$(\Co,X,\Io)$ in a Hilbert space $\cal H$ and an element
$\Omega$ of $\cal H$ with the property that
\be
{\cal F}(x,y)=\langle U(y)^*\Omega|U(x)^*\Omega\rangle
\ee
holds for all $x$ and $y$ in $X$.

\section*{Appendix B: Smeared-out field operators}

Here we discuss the relation between expressions (\ref{(2.6)})
and (\ref{emfcr}). 
The obvious relation between $\hat A(\psi)$
and $\hat A_\mu(q)$ is
\be
\hat A(\psi)=\int_{\Ro^4}\hbox{ d}q\,
f^\mu(q)\hat A_\mu(q).
\ee 
where the test functions $f_\mu$ correspond with
classical wavefunctions $\psi_\mu$ by (\ref{classwavefun}).
Similarly, let $\phi_\mu$ correspond with
test functions $g_\mu$.
Then one obtains, using the commutation relations (\ref{(2.6)})
and definition (\ref{(2.7)}) of $D_0(q)$,
\bea
\left[\hat
A(\psi),\hat A(\phi)\right] &=&\int_{\Ro^4}\hbox{ d}q\,
\int_{\Ro^4}\hbox{ d}q'\, f^\mu(q)g^\nu(q')
\left[\hat A_\mu(q),\hat A_\nu(q')\right]\cr
&=&-i\int_{\Ro^4}\hbox{ d}q\,
\int_{\Ro^4}\hbox{ d}q'\,
f^\mu(q)g_\mu(q')D_0(q-q')\cr
&=&-i\frac{1}{(2\pi)^3}\int_{\Ro^4}\hbox{ d}q\,
\int_{\Ro^4}\hbox{ d}q'\, f^\mu(q)g_\mu(q')\cr
& &\times
\int_{\Ro^3}\hbox{ d}{\bf k} \,\exp\left(i{\bf k}\cdot ({\bf
q}-{\bf q}')\right) {1\over |{\bf k}|}\sin((q_0-q'_0)|{\bf k}|)\cr
&=&-\frac{1}{(2\pi)^3}\int_{\Ro^3}\hbox{ d}{\bf k}\,\frac{1}{ 2|{\bf k}|} \int_{\Ro^4}\hbox{ d}q\,
f^\mu(q)\exp\left(i{\bf k}\cdot {\bf q}\right) \cr
& &\times \int_{\Ro^4}\hbox{ d}q'\,
g_\mu(q') \exp\left(-i{\bf k}\cdot {\bf q}'\right)\cr
& &\quad
\times \left(\exp\left(i(q_0-q'_0)|{\bf k}|\right)
-\exp\left(-i(q_0-q'_0)|{\bf k}|\right)\right).
\eea
Using the
definition (\ref{classwavefun}) of classical wave functions one
obtains
\bea
\left[\hat A(\psi),\hat A(\phi)\right]
&=&-\int_{\Ro^3}\hbox{ d}{\bf k}\,\frac{1}{2|{\bf k}|}
\left(\psi^\mu({\bf k})\overline{\phi_\mu({\bf k})}-
\overline{\psi^\mu({\bf k})}\phi_\mu({\bf k})\right).
\eea
The
latter implies (\ref{emfcr}).

\section*{Appendix C: Representative classical wave functions}

Here we show that each class of equivalent classical wave functions contains
a representative satisfying the radiation gauge.
Given the classical wave functions $\psi_\mu({\bf k})$, let
\be
\lambda({\bf k})
=\frac{1}{|{\bf k}|^2}\sum_{\alpha=1}^3{\bf k}_\alpha
\psi_\alpha({\bf k})
\ee
and
\bea
\phi_0({\bf k})&=&0, \qquad
\phi_\alpha({\bf k})
=\psi_\alpha({\bf k})-\lambda({\bf k}){\bf k}_\alpha,
\qquad\alpha=1,2,3.
\eea
Then one calculates
\bea
\sum_{\alpha=1}^3{\bf k}_\alpha \phi_\alpha({\bf k})
&=&\sum_{\alpha=1}^3{\bf k}_\alpha \psi_\alpha({\bf k})
-\lambda({\bf k})|{\bf k}|^2
=0.
\eea
Hence $\phi_\mu({\bf k})$ are classical wave functions satisfying
the radiation gauge.

Consider now an arbitrary set of Fourier coefficients $a_\mu({\bf k})$
satisfying (\ref{(1.14)}). Then one finds
\bea \langle a|\phi\rangle
&=&\int_{\Ro^3}\hbox{ d}{\bf k}\,\frac{1}{2|{\bf k}|}\,
\sum_{\alpha=1}^3\overline{a_\alpha({\bf k})}
\phi_\alpha({\bf k})\cr
&=&\int_{\Ro^3}\hbox{ d}{\bf k}\,\frac{1}{2|{\bf k}|}\,
\sum_{\alpha=1}^3\overline{a_\alpha({\bf k})} \psi_\alpha({\bf k})\cr
& &-\int_{\Ro^3}\hbox{ d}{\bf k}\,\frac{1}{2|{\bf k}|}\,
\lambda({\bf k})\sum_{\alpha=1}^3\overline{a_\alpha({\bf k})}
\,{\bf k}_\alpha.
\eea
Using (\ref{(1.14)}) this becomes
\bea
\langle a|\phi\rangle
&=&\int_{\Ro^3}\hbox{ d}{\bf k}\,\frac{1}{2|{\bf k}|}\,
\sum_{\alpha=1}^3\overline{a_\alpha({\bf k})}
\psi_\alpha({\bf k})\cr
& &-\frac{1}{2}\int_{\Ro^3}\hbox{ d}{\bf k}\,
\lambda({\bf k})\overline{a_0({\bf k})}\cr
&=&\int_{\Ro^3}\hbox{ d}{\bf k}\,\frac{1}{2|{\bf k}|}\,
\sum_{\alpha=1}^3\overline{a_\alpha({\bf k})}
\psi_\alpha({\bf k})\cr
& &-\int_{\Ro^3}\hbox{ d}{\bf k}\,
\frac{1}{2|{\bf k}|^2} \overline{a_0({\bf k})}
\sum_{\alpha=1}^3{\bf k}_\alpha\psi_\alpha({\bf k})\cr
&=&\langle a|\psi\rangle.
\eea
To obtain the latter,
(\ref{fermigauge}) has been used. This shows that $\phi$ and
$\psi$ are equivalent.

\section*{Appendix D: Existence of the vacuum state}

The difficult point to verify is positivity of the state. Note
that the symbols $\psi_j$ and $a_j$ below have each 4 components
$(\psi_j)_\mu$ respectively $(a_j)_\mu$. One verifies, using the
notation $b_j=a_j+i\psi_j$,
\bea
\sum_{jj'}\lambda_j\overline{\lambda_{j'}}
{\cal F}(a_j,\psi_j;a_{j'}\psi_{j'})
&=&\sum_{jj'}\lambda_j\overline{\lambda_{j'}}
\exp\left(-\frac{i}{2\eta}\Im\langle b_{j'}| b_j\rangle\right)\cr
& &\times
\exp\left(-\frac{1}{4\eta}\,\langle(b_{j'}-b_j|b_{j'}-b_j\rangle\right)\cr
&=&\sum_{j{j'}}\mu_j\overline{\mu_{j'}}
\exp\left(\frac{1}{2\eta}\langle b_{j}| b_{j'}\rangle\right)
\label{appctemp}
\eea
with
\be
\mu_j=\lambda_j\exp\left(-(1/4\eta)\,
\langle b_j|b_j\rangle\right).
\ee
Positivity of (\ref{appctemp}) follows by
means of Schur's lemma, provided we can show that the matrix with
elements $\langle b_{j}| b_{j'}\rangle$ is positive-definite. But
the latter is clear because of the positivity of the scalar
product, for which we now give a proof. Using the Lorenz condition
(\ref{(1.14)}, \ref{fermigauge}), one writes
\bea
\langle \phi|
\phi\rangle &=&\int_{\Ro^3}\hbox{ d}{\bf k}\,
\frac{1}{2|{\bf k}|} \left(-\overline{\phi_0({\bf k})}\phi_0({\bf k})
+\sum_{\alpha=1}^3\overline{\phi_\alpha({\bf k})}\phi_\alpha({\bf k})\right)\cr
&=&\int_{\Ro^3}\hbox{ d}{\bf k}\,\frac{1}{2|{\bf k}|^3}\,
\sum_{\alpha,\beta=1}^3\overline {\phi_\alpha({\bf k})}\,
\big[|{\bf k}|^2\delta_{\alpha,\beta}-{\bf k}_\alpha {\bf k}_\beta\big]
\,\phi_\beta({\bf k})\cr
&\ge&0.
\eea
The latter follows because the matrix
$|{\bf k}|^2\delta_{\alpha,\beta}-{\bf k}_\alpha {\bf k}_\beta$
is positive-definite.

\section*{Appendix E: Unitary representation of the group of shifts}

We show here that a unitary representation of the group of shifts $\Ro^4,+$
is determined by (\ref{unitshiftdef}).

Let us show that $U(x)$ is well-defined. Assume that
\be
\hat W(a^x,\psi^x)\Omega=\hat W(b^x,\phi^x)\Omega.
\ee
By taking the inner product with $\hat W(c,\chi)^*\Omega$ one obtains
\be
{\cal F}(a^x,\psi^x;c,\chi)={\cal F}(b^x,\phi^x;c,\chi).
\label{appDtemp}
\ee
Note now that $\langle a^x|\psi^x\rangle=\langle a|\psi\rangle$
so that
\be
{\cal F}(a^x,\psi^x;c,\chi)={\cal F}(a,\psi;c^{-x},\chi^{-x}).
\ee
Hence (\ref{appDtemp}) becomes
\be
{\cal F}(a,\psi;c^{-x},\chi^{-x})={\cal F}(b,\phi;c^{-x},\chi^{-x}).
\ee
This implies
\be
\langle \hat W(c^{-x},\chi^{-x})^*\Omega|(\hat W(a,\psi)^*-\hat W(b,\phi)^*)\Omega\rangle=0.
\ee
Since $c$ and $\chi$ are arbitrary, it follows that
$\hat W(a,\psi)^*\Omega=\hat W(b,\phi)^*\Omega$. This shows that $U(x)$ is well-defined.

It is now straightforward to show that $U(x)U(y)=U(x+y)$
and that $U(x)$ is isometric. Therefore, $U(x)$ is a unitary representation
of $\Ro^4,+$.



\newpage
\begin{thebibliography}{99}

\bibitem {CGH77} A.L. Carey, J.M. Gaffney, and C.A.Hurst,
{\sl A $C^*$-algebra formulation of the
quantization of the electromagnetic field,}
J. Math. Phys. {\bf 18}, 629-640 (1977).

\bibitem {JR80} J.M. Jauch and F. Rohrlich, {\sl The theory
of photons and electrons,} 2nd ed. (Springer-Verlag, 1980).

\bibitem{SG89} G. Scharf, {\sl Finite quantum electrodynamics}
(Springer-Verlag, 1989).

\bibitem{PD90} D. Petz, {\sl An Invitation to the Algebra of Canonical
Commutation Relations} (Leuven University Press, Leuven, 1990)

\bibitem {DFR94} S. Doplicher, K. Fredenhagen, J.E. Roberts,
{\sl Spacetime quantization induced by classical gravity,}
Phys. Lett. {\bf B331}, 39-44 (1994).

\bibitem {DFR95} S. Doplicher, K. Fredenhagen, J.E. Roberts,
{\sl The Quantum Structure of Spacetime at the Planck
Scale and Quantum Fields,} Commun. Math. Phys. {\bf 172}, 187-220 (1995)


\bibitem {NK00} J. Naudts, M. Kuna, {\sl Covariance systems,}
math-ph/0009031, J. Phys. A: Math. Gen. {\bf 34}, 9265-9280 (2001).

\bibitem {NK01} J. Naudts, M. Kuna, {\sl Model of a particle in
spacetime,} J. Phys. A: Math. Gen. {\bf 34}, 4227-4239 (2001).

\bibitem {NJ01} J. Naudts, {\sl Covariance approach to quantum theory,}
proceedings of the conference "Quantum Theory and Symmetries",
Krakow, July 2001.

\end{thebibliography}
\end{document}